\documentclass[prd,showpacs,preprintnumbers,nofootinbib]{revtex4}
\usepackage[dvips]{graphicx}
\usepackage{enumerate}
\usepackage{amsmath,amssymb}
\usepackage{mathrsfs}
\usepackage[dvips]{graphicx}
\usepackage{bm}

\begin{document}
\preprint{CHIBA-EP-213-v2, 2015}

\title{
Non-Abelian Stokes theorem for the Wilson loop operator in an arbitrary representation and its implication to quark confinement
}

\author{Ryutaro Matsudo}
\email{afca3071@chiba-u.jp}

\author{Kei-Ichi Kondo}
\email{kondok@faculty.chiba-u.jp}

\affiliation{Department of Physics,  
Graduate School of Science, 
Chiba University, Chiba 263-8522, Japan
}
\begin{abstract}
We give a gauge-independent definition of magnetic monopoles in the $SU(N)$ Yang-Mills theory through the Wilson loop operator.
For this purpose, we give an explicit proof of the Diakonov-Petrov version of  the non-Abelian Stokes theorem for the Wilson loop operator in an arbitrary representation of the $SU(N)$ gauge group to derive a new form for the non-Abelian Stokes theorem.
The new form is used to extract the magnetic-monopole contribution to the Wilson loop operator in a gauge-invariant way, which enables us to discuss confinement of quarks in any representation from the viewpoint of the dual superconductor vacuum. 
%The resulting formula can be applied to defining a gauge-invariant magnetic monopole 
%which enables one to discuss confinement of quarks in any representation from the viewpoint of the dual superconductor vacuum.
\end{abstract}


\maketitle
%%%%%%%%%%%%%%%%%%%%%%%%%%%%%%%%%%%%%%%%%%%%%%%%%

\section{Introduction}

%%%%%%%%%%%%%%%%%%%%%%%%%%%%%%%%%%%%%%%%%%%%%%%%%

The Wilson loop operator \cite{Wilson74} is the physical quantity of fundamental importance in gauge theories due to its gauge invariance. 
Indeed, quark confinement is judged by the area law of the vacuum expectation value of the Wilson loop operator, which is the so-called Wilson criterion for quark confinement. 
Recently, it has been shown \cite{KKSS15} that the \textbf{non-Abelian Stokes theorem (NAST)} for the Wilson loop operator is quite useful to understand quark confinement based on the dual superconductor picture \cite{dualsuper}. 
Here the NAST for the Wilson loop operator refers to the alternative expression  in which the line integral defining the original Wilson loop operator is replaced  by the surface integral. 
In particular, we want to obtain the NAST  which eliminates the {path ordering}. 
Such a version of the NAST was indeed derived for the first time by Diakonov and Petrov for the $SU(2)$ Wilson loop operator based on a specific method in \cite{DP89} (See also \cite{DP96}).
%[Diakonov and Petrov, Phys.~Lett. B224, 131 (1989)].
Later, it was recognized that the Diakonov-Petrov version of the NAST can be derived as a path-integral representation using the \textbf{coherent state of the Lie group} in a unified way \cite{KondoIV,KT00b,KT00,Kondo08}.
The NAST is rederived based on the $SU(2)$ coherent state in \cite{KondoIV}.
%[Kondo, hep-th/9805153, Phys.~Rev. D58, 105016 (1998)]. 
In a similar way, the NAST has been extended into the gauge group $SU(3)$ in \cite{KT00b} and
%[Kondo \& Taira, hep-th/9906129, Mod.~Phys.~Lett. A15, 367 (2000)].
  $SU(N)$ in \cite{KT00,Kondo08} 
%[Kondo \& Taira, hep-th/9911242, Prog.~Theor.~Phys. 104, 1189 (2000)]
%and 
%[Kondo, hep-th/0009152]. 
%The presentation for $SU(N)$  of this chapter is based on the paper  \cite{Kondo08}.
%[Kondo, arXiv:0801.1274, Phys.~Rev.D77, 085029 (2008)]. 
to discuss the quarks in the \textbf{fundamental representation} \cite{Kondo99Lattice99,Kondo00,Kondo08b}.
See \cite{KKSS15} for a review. 
There exist other versions of the NAST, see \cite{HU99,Halpern79,Bralic80,Arefeva80,Simonov89,Lunev97,HM97}. 

Let $\mathscr{A}$ be the Lie algebra valued \textbf{connection one-form} for the gauge group $G=SU(N)$: 
\begin{equation}
 \mathscr{A}(x) :=  \mathscr{A}_\mu(x)  dx^\mu  = \mathscr{A}_\mu^A(x) T_A dx^\mu \in \mathscr{G}=Lie(G)
 \quad (A=1,..., {\rm dim}G)
  ,
\end{equation}
where $T_A$ is the generator of the Lie algebra $\mathscr{G}=su(N)$ of the group $G=SU(N)$ and ${\rm dim}G$ is the dimension of the group $G$, i.e., ${\rm dim}G=N^2-1$ for $G=SU(N)$. 
In what follows, the summation over the repeated indices should be understood unless otherwise stated. 
For a given loop, i.e.,  a closed path $C$,   the  \textbf{Wilson loop operator} $W_{\rm C}[\mathscr{A}]$ in the representation $R$ is defined by
\begin{equation}
 W_C[\mathscr{A}] := \mathcal{N}^{-1} {\rm tr}_{R} \left\{ \mathscr{P} \exp \left[ -ig_{{}_{\rm YM}} \oint_C \mathscr{A} \right] \right\} ,
 \quad  \mathcal{N}:=d_R = {\rm tr}_R({\bf 1}) ,
\end{equation}
where $\mathscr{P}$ denotes the \textbf{path ordering} and  the normalization factor $\mathcal{N}$ is equal to the dimension $d_R$ of the representation $R$, to which the probe of the Wilson loop belongs,   
ensuring $W_C[0]=1$.
We introduce the Yang-Mills coupling constant $g_{{}_{\rm YM}}$  for later convenience, although this can be absorbed by scaling the field $\mathscr{A}^\prime=g_{{}_{\rm YM}}\mathscr{A}$.

For the gauge group $SU(2)$, for instance, any representation is characterized by  a single index  $J=\frac12, 1, \frac32, 2, \frac52, \cdots$. 
In fact, the Wilson loop operator in the representation $J$ of $SU(2)$ is rewritten into the surface-integral form \cite{DP89,KondoIV}:
\begin{align}
  W_C[\mathscr{A}]  
 =&  \int  [d \mu(g)]_\Sigma \exp \left\{ -i  g_{{}_{\rm YM}} J \int_{\Sigma: \partial \Sigma=C} dS^{\mu\nu} f_{\mu\nu}^g \right\} , 
% \ \text{no path-ordering}
  \nonumber\\
  f_{\mu\nu}^g(x) =& \partial_\mu [ {n}^A(x)   \mathscr{A}^A_\nu(x) ] -  \partial_\nu [{n}^A(x)  \mathscr{A}^A_\mu(x)  ]   
%\nonumber\\ &
  - g_{{}_{\rm YM}}^{-1} \epsilon^{ABC}  {n}^A(x)  \partial_\mu  {n}^B(x)  \partial_\nu  {n}^C(x) 
%  f_{\mu\nu}(x) =& \partial_\mu [\bm{n}(x) \cdot \mathscr{A}_\nu^A(x)] -  \partial_\nu [\bm{n}(x) \cdot \mathscr{A}_\mu(x)]   
%  - g^{-1} \bm{n}(x) \cdot [\partial_\mu \bm{n}(x) \times \partial_\nu \bm{n}(x) ]
 ,
  \nonumber\\
%  \bm{n}(x) &:= 
 n^A(x) \sigma^A   =& g(x)  \sigma^3 g^\dagger(x) , \ g(x) \in SU(2) \ (A,B,C \in \{ 1,2,3 \}) ,
\end{align}
where $\sigma^A$ ($A=1,2,3$) are the Pauli matrices with $\sigma^3$ being the diagonal matrix, $g$ is an $SU(2)$ group element and $ [d \mu(g)]_\Sigma$ is the product of an invariant measure on $SU(2)/U(1)$ over $\Sigma$: 
\begin{align}
 [d \mu(g)]_\Sigma :=\prod_{x \in \Sigma}d\mu(\bm{n}(x)) ,
  \ 
  d\mu(\bm{n}(x)) = \frac{2J+1}{4\pi} \delta( {n}^A(x) {n}^A(x)-1) d^3 \bm{n}(x) .  
\end{align}
 
%%%%%%%%%%%%%%%%%%%%%%%%%%%%%%%%%%%%%%%%%%%%%%%%%

The purpose of this paper is to extend the Diakonov-Petrov version of the NAST for the Wilson loop operator to an arbitrary representation of the $SU(N)$ group ($N \ge 3$) to derive a new form for the NAST, which enables one to define a gauge-invariant magnetic monopole in the Yang-Mills theory and to extract the magnetic-monopole contribution to the Wilson loop operator in a gauge-invariant way. 
The new form of the NAST has been obtained already for the fundamental representation of $SU(N)$ ($N \ge 3$) in \cite{Kondo08}.
The new form is useful to discuss quark confinement in an arbitrary representation from the viewpoint of the dual superconductor picture. 
The relevance of the Wilson loop to quark confinement can be observed by calculating the magnetic monopole current $k$, whose definition is proposed in this paper.
In fact, one of the authors and his collaborators have used the new form of the NAST  to calculate the average of the Wilson loop operator for $SU(2)$ and  $SU(3)$ in the fundamental representation using the numerical simulations on a lattice. 
Through the simulations, they have examined the dual superconductivity picture for quark confinement.
See chapter 9 of \cite{KKSS15}. 
The new form of the NAST will be used to extend the preceding works to any representation of $SU(3)$ in subsequent works.

Last but not least we must mention the facts that an original form of the NAST for arbitrary representation of the $SU(N)$ group ($N \ge 3$) was already announced in the second paper of Ref.\cite{DP96} and that the same form for the NAST has been derived in an independent way specifically for the fundamental representation of the $SU(N)$ gauge group in \cite{Lunev97}. 
However, the formula given there is not appropriate for our purpose stated above. 
Although the formula given originally in the second paper of \cite{DP96} is correct, indeed,  nontrivial (mathematical) works are required to derive the new form from it. 
Moreover, to the best of our knowledge, there are no available proofs of the NAST for any representation in the published literature. Therefore, we  give  an explicit proof of the NAST as a preliminary step toward our purpose.

%The purpose of this paper is to extend the NAST for the Wilson loop operator to the arbitrary representation of $SU(N)$ group ($N \ge 3$).
%Moreover, we give a gauge-invariant definition of the magnetic monopole from the Wilson loop operator, which is useful to discuss quark confinement in arbitrary representation from the viewpoint of the dual superconductor picture. 

%In ref.[13] eq. (3) has been rederived in an independent way specifically for the fundamental representation of the SU(N) gauge group.

%for fundamental representation of SU(N) coset is SU(N)/SU(N-1)/U(1) * CPN*1. Correspondingly in the last case coset can be parametrized by unit complex vector u.The corresponding NAST was first published in [18], see also [5] .

\section{non-Abelian Stokes theorem for the Wilson loop operator}

%%%%%%%%%%%%%%%%%%%%%%%%%%%%%%%%%%%%%%%%%%%%%%%%%
Let $\mathscr{A}^{g}(x) =  \mathscr{A}^{g}_\mu(x) dx^\mu$ be the gauge transformation of the Yang-Mills gauge field $\mathscr{A}(x)$ by the group element $g \in G$: 
\begin{align}
\mathscr{A}^{g}(x)  :=  {g}(x)^\dagger \mathscr{A}(x) {g}(x)
+ ig_{{}_{\rm YM}}^{-1} {g}(x)^\dagger d {g}(x) .
\end{align}
Using a \textbf{reference state} $\left| {\Lambda} \right>$, we define the one-form $A^{g}(x) = A^{g}_\mu(x) dx^\mu$ from the  Lie algebra valued one-form $\mathscr{A}^{g}(x) = \mathscr{A}^{g}_\mu(x) dx^\mu$ by 
\begin{align}
A^g(x)  :=& \langle \Lambda | \mathscr{A}^{g}(x)      |\Lambda \rangle  
\quad
\text{or} \quad
%\mathscr{A}^{g}(x)  :=  {g}(x)^\dagger \mathscr{A}(x) {g}(x)
%+ ig_{{}_{\rm YM}}^{-1} {g}(x)^\dagger d {g}(x)  ,
%\nonumber\\
   A^{g}_\mu(x) =  \langle \Lambda | \mathscr{A}_\mu^{g} (x)    |\Lambda \rangle   .
%  A^{g} =  A^{g}_\mu(x) dx^\mu  =  \langle \Lambda | \mathscr{A}_\mu^{g} (x)    |\Lambda \rangle  dx^\mu .
%\quad 
%\mathscr{A}^{g} =  \mathscr{A}^{g}_\mu(x) dx^\mu ,  
%\mathscr{A}^{g}_\mu(x)  :=  {g}(x)^\dagger \mathscr{A}_\mu(x) {g}(x)
%+ ig_{{}_{\rm YM}}^{-1} {g}(x)^\dagger \partial_\mu {g}(x)  ,
%\\
%  \mathscr{A}_\mu^{g}(x) :=&  {g}(x)^\dagger \mathscr{A}_\mu(x) {g}(x) 
%+ ig_{{}_{\rm YM}}^{-1} {g}(x)^\dagger \partial_\mu {g}(x)  .
\label{C29-pre-NAST0}
\end{align}
Then it is shown \cite{KondoIV,KT00} that the Wilson loop operator has a path-integral representation,
%[Exercise-2] \marginpar{Ex-2}
%Verify that $\mathscr{A}^{g}$ defined by (\ref{C29-def-Ag0}) is written in the form (\ref{C29-Ag2}). 
\begin{align}
 W_C[\mathscr{A}] 
 =&  \int [d\mu({g})]_C \exp \left( 
-ig_{{}_{\rm YM}}  \oint_C  A^g \right) , 
%\quad 
%[d\mu({g})]_C =  \prod_{x \in C}  d\mu({g}(x)) 
%:= \lim_{N \rightarrow \infty, \epsilon \rightarrow 0} \prod_{n=0}^{N-1}  d\mu({g}_n) ,
%\nonumber\\
% A^g  :=& \langle \Lambda | \mathscr{A}^{g}      |\Lambda \rangle ,
%\nonumber\\
% \mathscr{A}^{g}(x) :=&  \mathscr{A}^{g}_\mu(x) dx^\mu , \ \mathscr{A}^{g}_\mu(x)  :=  {g}(x)^\dagger \mathscr{A}_\mu(x) {g}(x)
%+ ig_{{}_{\rm YM}}^{-1} {g}(x)^\dagger \partial_\mu {g}(x)  ,
\label{C29-Ag2}
\end{align}
where 
$[d\mu({g})]_C$ is the product of the invariant integration measure $d\mu({g}(x))$ at each   point $x$ on the loop $C$:
\begin{align}
  [d\mu({g})]_C =  \prod_{x \in C}  d\mu({g}(x)) .
%:=  \lim_{N \rightarrow \infty, \epsilon \rightarrow 0} \prod_{n=0}^{N-1}  d\mu({g}_n) ,
\end{align}
%and $d$ denotes the exterior derivative:
%\begin{equation}
% d = ds \frac{d}{ds} = ds \frac{dx^\mu}{ds} \frac{\partial}{\partial x^\mu} =    dx^\mu \frac{\partial}{\partial x^\mu} = dx^\mu \partial_\mu  .
%\end{equation}

%%%%%%%%%%%%%%%%%%%%% figures %%%%%%%%%%%%%%%%%%%%%%%%%%%
\begin{figure}[tbp]
\begin{center}
\includegraphics[height=3.0cm]{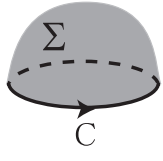}
\end{center} 
\vskip -0.5cm
\caption[]{
A closed loop $C$ for defining the Wilson loop operator and the surface $\Sigma$ whose boundary is given by the loop $C$. 
}
\label{C29-fig:W_loop-Sigma}
\end{figure}
%%%%%%%%%%%%%%%%%%%%% figures %%%%%%%%%%%%%%%%%%%%%%%%%%%

Now the argument of the exponential is an Abelian quantity, since $A^{g}_\mu$ is no longer a matrix, just a number. 
Therefore, we can apply  the (usual) \textbf{Stokes theorem},
\begin{equation}
 \oint_{C=\partial \Sigma} \omega = \int_\Sigma d \omega   ,
\end{equation}
to replace the line integral along the closed loop $C$ to the surface integral over the surface $\Sigma$ bounded by $C$.
See Fig.~\ref{C29-fig:W_loop-Sigma}.
Thus we obtain a NAST:
\begin{equation}
 W_C[\mathscr{A}]  =\int [d\mu(g)]_{\Sigma}
\exp \left[ -ig_{{}_{\rm YM}} \int_{\Sigma: \partial \Sigma=C} F^g  \right] , %\quad
%[d\mu(g)]_{\Sigma} :=\prod_{x \in \Sigma: \partial \Sigma=C}   d\mu(g(x))  ,
\label{C29-NAST1}
\end{equation}
where  the $F^g$ is the curvature two-form  defined by  
\begin{equation}
 F^g  =dA^g  = \frac12 F^g_{\mu\nu}(x) dx^\mu \wedge dx^\nu , \quad
 F^g_{\mu\nu}(x)
:=  \partial_\mu A_\nu^{g}(x) - \partial_\nu A_\mu^{g}(x) ,
\end{equation}
and the integration measure on the loop $C$ is replaced by the integration measure on the surface $\Sigma$,
\begin{equation}
 [d\mu(g)]_{\Sigma} :=\prod_{x \in \Sigma: \partial \Sigma=C}   d\mu(g(x))  ,
\end{equation}
by inserting additional integral measures,  $1=\int d\mu(g(x))$ for $x \in \Sigma - C$.

The field strength $F^g_{\mu\nu}$ is calculated as
\begin{align}
  F^g_{\mu\nu}
=& \partial_\mu A_\nu^{g} - \partial_\nu A_\mu^{g} 
\nonumber\\ 
=&  \partial_\mu \langle \Lambda | \mathscr{A}_\nu^{g} |\Lambda \rangle - \partial_\nu \langle \Lambda | \mathscr{A}_\mu^{g} |\Lambda \rangle
\nonumber\\ 
=&  \partial_\mu \langle \Lambda | g^\dagger \mathscr{A}_\nu g |\Lambda \rangle - \partial_\nu \langle \Lambda | g^\dagger \mathscr{A}_\mu g|\Lambda \rangle
%\nonumber\\&
+ ig_{{}_{\rm YM}}^{-1} \partial_\mu \langle \Lambda | g^\dagger \partial_\nu g |\Lambda \rangle 
- ig_{{}_{\rm YM}}^{-1} \partial_\nu \langle \Lambda | g^\dagger \partial_\mu g|\Lambda \rangle 
\nonumber\\
=& \partial_\mu \langle \Lambda | g^\dagger \mathscr{A}_\nu g |\Lambda \rangle - \partial_\nu \langle \Lambda | g^\dagger \mathscr{A}_\mu g|\Lambda \rangle
%\nonumber\\&
+ ig_{{}_{\rm YM}}^{-1}  \langle \Lambda | ( \partial_\mu g^\dagger \partial_\nu g  -  \partial_\nu g^\dagger \partial_\mu g) |\Lambda \rangle
+ ig_{{}_{\rm YM}}^{-1}  \langle \Lambda | g^\dagger [ \partial_\mu, \partial_\nu ] g |\Lambda \rangle  
\nonumber\\
=& \partial_\mu \langle \Lambda | g^\dagger \mathscr{A}_\nu g |\Lambda \rangle - \partial_\nu \langle \Lambda | g^\dagger \mathscr{A}_\mu g|\Lambda \rangle
%\nonumber\\&
+ ig_{{}_{\rm YM}}   \langle \Lambda | g^\dagger [\Omega_\mu , \Omega_\nu] g |\Lambda \rangle
+ ig_{{}_{\rm YM}}  \langle \Lambda | g^\dagger [ \partial_\mu, \partial_\nu ] g |\Lambda \rangle  
,
\end{align}
where we have introduced 
\begin{align}
 \Omega(x) :=   ig_{{}_{\rm YM}}^{-1} {g}(x) d {g}^\dagger(x) .
\end{align}

We define the Lie algebra valued   field $m(x)$ which we call the \textbf{precolor  (direction)  field} by
\begin{align}
 \bm{m}(x) := \left< {\Lambda} |  g^\dagger(x) T_A g(x) | {\Lambda} \right> T_A 
= m^A(x) T_A ,
 \quad 
 {m}^A(x) = \left< {\Lambda} |  g^\dagger(x) T_A g(x) | {\Lambda} \right> .
\label{m2b} 
\end{align}
For a Lie algebra valued operator $\mathscr{O}(x)=\mathscr{O}^A(x) T_A$, we obtain the relation:
\begin{align}
 \langle \Lambda | g^\dagger(x) \mathscr{O}(x)  g(x) |\Lambda \rangle  
= \langle \Lambda | g^\dagger(x) T_A g(x) |\Lambda \rangle \mathscr{O}^A(x) 
= m^A(x) \mathscr{O}^A(x)  
= \kappa  {\rm tr}(\bm{m}(x)\mathscr{O}(x) )
 ,
\end{align}
where we adopted the normalization for the generator:
\begin{align}
 {\rm tr}(T_A T_B) = \kappa^{-1}  \delta_{AB} 
 .
\end{align}
Therefore, the field strength $F^g_{\mu\nu}$ is written as
\begin{align}
  F^g_{\mu\nu}(x) 
=& \kappa \{ \partial_\mu  {\rm tr}(\bm{m}(x)\mathscr{A}_\nu(x) ) - \partial_\nu  {\rm tr}(\bm{m}(x)\mathscr{A}_\mu(x) )
+ ig_{{}_{\rm YM}}   {\rm tr}(\bm{m}(x)[\Omega_\mu (x) , \Omega_\nu (x)]) \} 
\nonumber\\&
+ ig_{{}_{\rm YM}}  \langle \Lambda | g^\dagger(x) [ \partial_\mu, \partial_\nu ] g(x) |\Lambda \rangle  
 .
\label{Fg2}
\end{align} 
Notice that the final term is not gauge invariant and disappears finally after the integration with respect to the gauge-invariant measure $d\mu(g)$. Therefore, it is omitted in what follows.

%%%%%%%%%%%%%%%%%%%%%%%%%%%%%%%%%%%%%%%%%%%%%%%%%

\section{Color direction field}

%%%%%%%%%%%%%%%%%%%%%%%%%%%%%%%%%%%%%%%%%%%%%%%%%
As a reference state $\left| {\Lambda} \right>$, we can choose the \textbf{highest-weight state} defined by the (normalized) common eigenvector of the generators $H_1, H_2, \cdots, H_r$ in the \textbf{Cartan subalgebra} with the  eigenvalues $\Lambda_1, \Lambda_2, \cdots, \Lambda_r$:
\begin{equation}
H_j \left|   {\Lambda} \right> 
=\Lambda_j \left|   {\Lambda} \right> 
 \quad ( j=1, \cdots, r ) 
   ,
   \label{C28-eigen3}
\end{equation}
where $r$ is the rank of $G$, i.e., $r:={\rm rank}G=N-1$. 
Then we have 
\begin{align}
 \left< {\Lambda} | H_j | {\Lambda} \right> 
=  \Lambda_j \left< {\Lambda} | {\Lambda} \right> 
= \Lambda_j 
 \quad ( j=1, \cdots, r ) 
   ,
   \label{C28-eigen4}
\end{align}
by taking into account the normalization $\left< {\Lambda} | {\Lambda} \right>=1$.

Let 
$\mathcal{R}_+$ $(\mathcal{R}_-)$ be a subsystem of \textbf{positive (negative) roots}.%
\footnote{
The root vector is defined to be the weight vector of the adjoint representation. 
A weight $\vec{\nu}_j$ is called positive if its {\it last} nonzero component is positive. With this definition, the weights satisfy 
$
 \nu^1 > \nu^2 > \cdots > \nu^N  
%\label{order}
$.
}
Then the highest-weight state satisfies the following  properties:
\begin{enumerate}
%\item[(i)] 
%$|\Lambda \rangle$  is mapped into itself by all diagonal operators $H_j$,
%$
% H_{j} |\Lambda \rangle = \Lambda_j |\Lambda \rangle ;
%$

\item[(i)]
 $|\Lambda \rangle$  is
annihilated by all the (off-diagonal) \textbf{shift-up operators} $E_{\alpha}$ with
$\alpha \in \mathcal{R}_+$: 
%$
% E_{\alpha} |\Lambda \rangle = 0 \ (\alpha \in R_+)  
%$,
\begin{equation}
% H_j | \Lambda \rangle = \Lambda_j | \Lambda \rangle
% , \quad
 E_\alpha | \Lambda \rangle = 0 \quad (\alpha \in \mathcal{R}_+)
 , 
\end{equation}

\item[(ii)] 
$|\Lambda \rangle$  is
annihilated by some \textbf{shift-down operators} $E_{\alpha}$ with
$\alpha \in \mathcal{R}_-$, not by other $E_{\beta}$ with $\beta \in \mathcal{R}_-$:
\begin{equation}
 E_{\alpha} |\Lambda \rangle = 0 \ ({\rm some~} \alpha \in \mathcal{R}_-) ;
 \quad
 E_{\beta} |\Lambda \rangle = |\Lambda+\beta \rangle 
 \ ({\rm some~} \beta \in \mathcal{R}_-) .
 \label{C28-ii}
\end{equation}

\end{enumerate}

The adjoint rotation of a generator $T_A$ can be written as a linear combination of the generators $\{ T_A \}$:
\begin{align}
 g^\dagger(x) T_A g(x)  =&  R_{AB}(x) T_B 
 ,
 \label{gTgd2}
\end{align}
since 
$g^\dagger(x) T_A g(x)$ is written by using the commutator repeatedly:
\begin{align}
g^\dagger(x) T_A g(x)
=e^{iY} T_A e^{-iY}
=T_A + [iY, T_A] + \frac12 [iY, [iY, T_A]] + ..., \quad Y := \theta^B T_B ,
\end{align}
and the commutator is closed $[T_A, T_B]=if_{ABC}T_C$ with the structure constant $f_{ABC}$.
Hence, the precolor field (\ref{m2b}) is written as
\begin{align}
 \bm{m}(x) 
%=  \left< {\Lambda} | T_B | {\Lambda} \right> g(x) T_B g^\dagger(x) 
= R_{AB}(x) \left< {\Lambda} | T_B | {\Lambda} \right>   T_A
,  \quad 
 {m}^A(x) 
%=  \left< {\Lambda} | T_B | {\Lambda} \right> g(x) T_B g^\dagger(x) 
= R_{AB}(x)  \left< {\Lambda} | T_B | {\Lambda} \right>  
 .
\label{m3b}
\end{align}
%The adjoint rotation of a generator $T_B$ can be written as a linear combination of the generators $\{ T_A \}$:
%\begin{align}
% g(x) T_B g^\dagger(x)  =&  T_A R_{AB}(x) 
% .
% \label{gTgd2b}
%\end{align}
Multiplying $g(x)$ from the left and $g^\dagger(x)$ from the right, on the other hand, (\ref{gTgd2}) yields 
\begin{align}
  T_A   =&  R_{AC}(x) g(x) T_C g^\dagger(x)  
 ,
\end{align}
which is cast after multiplying $R_{AB}(x)$  into the form:
\begin{align}
  R_{AB}(x) T_A   
=&   R_{AB}(x) R_{AC}(x)  g(x) T_C g^\dagger(x) 
\nonumber\\
=&  (R^t)_{BA}(x) R_{AC}(x)  g(x) T_C g^\dagger(x) 
\nonumber\\
=&  \bm{1}_{BC}  g(x) T_C g^\dagger(x) 
\nonumber\\
=&  g(x) T_B g^\dagger(x)  
 ,
\label{TR2}
\end{align}
where we have used the fact that the matrix $R$ is a real-valued  $R_{AB}^*=R_{AB}$ and unitary $R^\dagger R=R R^\dagger=\bm{1}$, in other words, $R$ is an orthogonal matrix satisfying $R^tR=RR^t=\bm{1}$ for the transposed matrix $R^t$ of $R$, 
because the structure constant is real-valued.

By substituting (\ref{TR2}) into (\ref{m3b}), the precolor field is written as
\begin{align}
 \bm{m}(x) 
%=  m^A(x) T_A
%=&  \left< {\Lambda} | T_B | {\Lambda} \right>  T_A R_{AB}(x) 
%\nonumber\\
=& \left< {\Lambda} | T_B | {\Lambda} \right> g(x) T_B g^\dagger(x) 
\nonumber\\
=& \left< {\Lambda} | H_j | {\Lambda} \right> g(x) H_j g^\dagger(x) 
\nonumber\\
=& \Lambda_j  g(x) H_j g^\dagger(x) 
 ,
\label{m1b}
\end{align}
where we have used in the second equality  the fact that the generators $T_A$ other than the Cartan generators $H_j$, i.e., the shift-up and shift-down generators $E_{\alpha}$ in the Cartan basis have the property:
\begin{align}
  \left< {\Lambda} | E_{\alpha} | {\Lambda} \right> = 0 
 ,
\end{align}
since $E_{\alpha} \left| {\Lambda} \right> = 0$ or $E_{\beta} \left| {\Lambda} \right>$ is the eigenvector with the eigenvalue ${\Lambda}+{\beta}$ and obeys
\begin{align}
  \left< {\Lambda} | E_{\beta} | {\Lambda} \right> = N_{\beta, \Lambda} \left< {\Lambda} |  {\Lambda}+{\beta} \right> = 0 
 ,
\end{align}
because the eigenvectors with the different eigenvalues are orthogonal 
$\left< {\Lambda} |  {\Lambda}^\prime \right>=0$ for $\Lambda \ne \Lambda^\prime$. 
We have used (\ref{C28-eigen4}) in the last equality.

%Therefore, the result (\ref{m2b}) is obtained by substituting (\ref{TRb}) into (\ref{m3b}).

We introduce $r$ Lie algebra valued fields $\bm{n}_j(x)$ defined by
\begin{align}
 \bm{n}_j(x) := g(x) H_j g^\dagger(x) = n^A_j(x) T_A \quad (j=1,..., r) .
\end{align}
Then we arrived at the important relation:
\begin{align}
 \bm{m}(x) :=  \Lambda_j \bm{n}_j(x) \in \mathscr{G}=Lie(G)=su(N),
\quad m^A(x) :=  \Lambda_j n^A_j(x) .
\end{align}
%We introduce the Lie-algebra value field $\bm{m}(x)$ which we call the precolor field by 
%\begin{align}
% \bm{m}(x) := g(x) \Lambda_j H_j g^\dagger(x) = m^A(x) T_A \in \mathscr{G}=Lie(G)=su(N), \quad g \in G = SU(N).
%\end{align}
Notice that (\ref{gTgd2}) is determined by the commutation relation alone and, hence, $R_{AB}$ does not depend on the representation adopted. 
Therefore, $n^A_j(x)$ does not depend on the representation 
\begin{align}
  n^A_j(x) = R_{Aj}(x) ,
\end{align}
and we can use the fundamental representation to calculate $n^A_j(x)$ and to calculate the precolor field $m^A(x)$. 
\begin{align}
 {m}^A(x) = \left< {\Lambda} |  g^\dagger(x) T_A g(x) | {\Lambda} \right> 
=  \Lambda_j n^A_j(x) 
= \Lambda_j g(x) H_j g^\dagger(x)  .
\end{align}

%%%%%%%%%%%%%%%%%%%%%%%%%%%%%%%%%%%%%%%%%%%%%%%%%

\section{Derivation}

%%%%%%%%%%%%%%%%%%%%%%%%%%%%%%%%%%%%%%%%%%%%%%%%

We define $\mathscr{B}_\mu(x)$ by
\begin{align}
  \mathscr{B}_\mu(x) := ig_{{}_{\rm YM}}^{-1} 
%\sum_{j=1}^{r} 
[ \bm{n}_j(x), \partial_\mu  \bm{n}_j(x) ]  
%   =   g^{-1}  (\partial_\mu  \bm{n}_j(x) \times \bm{n}_j(x) )
 .
\label{C27-B}
\end{align}
In what follows, the summation over $j$ should be understood.
Then it satisfies the relation:
\begin{align}
 ig_{{}_{\rm YM}} [\mathscr{B}_\mu(x), \bm{n}_j(x)]  %\mathscr{D}_\mu[\mathscr{B}] \bm{n}_j(x) 
= \partial_\mu \bm{n}_j(x)  
\quad (j=1,2, \cdots, r)  .
\label{dnB}
\end{align}
The relation (\ref{dnB}) is derived in Appendix~\ref{section:derivation-eq}.
Hence we obtain a relation for the precolor field $\bm{m} (x)=\Lambda_j \bm{n}_j(x)$:
\begin{align}
 \partial_\mu \bm{m} (x)  = ig_{{}_{\rm YM}} [\mathscr{B}_\mu(x), \bm{m} (x)] . %\mathscr{D}_\mu[\mathscr{B}] \bm{n}_j(x) 
 \label{Brel}
\end{align}

On the other hand, we find  
\begin{align}
\partial_\mu \bm{n}_j(x) = ig_{{}_{\rm YM}} [ \Omega_\mu , \bm{n}_j(x) ] .
\label{del-n2}
\end{align}
This relation follows from  
\begin{align}
   \partial_\mu {\bm{n}_j} 
 =  \partial_\mu (g H_j g^\dagger) 
& = \partial_\mu  g g^\dagger g H_j g^\dagger  +   g  H_j g^\dagger g \partial_\mu g^\dagger 
%\quad (    g^\dagger g =1)
\nonumber\\ &
= -   g \partial_\mu g^\dagger g H_j g^\dagger  +   g  H_j g^\dagger g \partial_\mu g^\dagger 
%\quad (  \partial_\mu  g g^\dagger =-g \partial_\mu g^\dagger)
\nonumber\\ &
=  -[ g \partial_\mu g^\dagger  , g H_j g^\dagger ] 
%\nonumber\\ &
%= -[ g \partial_\mu g^\dagger  , \tilde{\bm{n}} ] 
\nonumber\\ &
=  ig_{{}_{\rm YM}}[ \Omega_\mu ,{\bm{n}_j}] ,
%\nonumber\\&
%\quad \Longrightarrow \quad 
%\partial_\mu \bm{n} = ig_{{}_{\rm YM}} [ \Omega_\mu , \bm{n} ] ,
\end{align}
where we have used $g^\dagger g=\bm{1}=g g^\dagger$ in the second equality and $\partial_\mu  g g^\dagger =-g \partial_\mu g^\dagger$ following from $\partial_\mu(g g^\dagger )=0$ in the third equality.
Therefore, we obtain another relation for the precolor field $\bm{m} (x)=\Lambda_j \bm{n}_j(x)$:
\begin{align}
\partial_\mu \bm{m}(x)  = ig_{{}_{\rm YM}} [ \Omega_\mu(x) , \bm{m}(x)  ] ,
\label{del-n2b}
\end{align}
Combining (\ref{Brel}) and (\ref{del-n2b}), we conclude
\begin{align}
 [ \Omega_\mu(x) , \bm{m}(x)  ] =  [\mathscr{B}_\mu(x), \bm{m} (x)]    .
\label{dmm}
\end{align}
The relation (\ref{dmm}) is used to write the third term in $F^g_{\mu\nu}(x)$ as 
\begin{align}
  ig_{{}_{\rm YM}}   {\rm tr}(\bm{m} [\Omega_\mu , \Omega_\nu ]) 
%=&  
%  ig_{{}_{\rm YM}}   {\rm tr}([\bm{m},\Omega_\mu ] \Omega_\nu  ) 
%\nonumber\\ 
%=& ig_{{}_{\rm YM}}   {\rm tr}(\Omega_\nu [\bm{m}, \mathscr{B}_\mu ]   )  
%\nonumber\\ 
%=& ig_{{}_{\rm YM}}   {\rm tr}([\Omega_\nu , \bm{m}] \mathscr{B}_\mu )  
%\nonumber\\ 
%=& ig_{{}_{\rm YM}}   {\rm tr}([\mathscr{B}_\nu , \bm{m}] \mathscr{B}_\mu )  
%\nonumber\\ 
%=&  {\rm tr}(\partial_\nu \bm{m} \mathscr{B}_\mu )  
%\nonumber\\ 
%=& ig_{{}_{\rm YM}}^{-1}  {\rm tr}(\partial_\nu \bm{m} [ \bm{n}_j , \partial_\mu  \bm{n}_j ])  
%\nonumber\\ 
%=& ig_{{}_{\rm YM}}^{-1}  {\rm tr}( [\partial_\nu \bm{m} ,   \bm{n}_j ] \partial_\mu  \bm{n}_j  )  
%\nonumber\\ 
%=& ig_{{}_{\rm YM}}^{-1}  {\rm tr}( [ \partial_\nu  \bm{n}_j , \bm{m} ] \partial_\mu  \bm{n}_j  )  
%\nonumber\\ 
=  ig_{{}_{\rm YM}}^{-1}  {\rm tr}( [\partial_\mu  \bm{n}_j , \partial_\nu  \bm{n}_j ] \bm{m} )  
,
\label{Fg3}
\end{align} 
%where we have used ${\rm tr}\{ A [ B, C ]\}={\rm tr}\{ [A , B]  C  \}={\rm tr}\{ B [ C , A ]\}={\rm tr}\{ C [ A , B ]\}$ due to the cyclicity of the trace in the first, third, seventh equalities, and the relation
%$[ \partial_\nu \bm{m} , \bm{n}_j ] = [\partial_\nu \bm{n}_j , \bm{m} ]$ 
%which is derived from
%$\partial_\nu [\bm{m} , \bm{n}_j]=0$
%and the commutativity
%$[ \bm{m} , \bm{n}_j ]= 0$ in the eight equality.
%\begin{align}
%  [ \bm{m} , \bm{n}_j ]= 0 \Longrightarrow \partial_\nu [\bm{m} , \bm{n}_j]=0 \Longrightarrow [ \partial_\nu \bm{m} , \bm{n}_j ] = [\partial_\nu \bm{n}_j , \bm{m} ]. 
%\end{align}
The relation (\ref{Fg3}) is also derived in Appendix~\ref{section:derivation-eq}.
Therefore, the field strength $F^g_{\mu\nu}$ is written as
\begin{align}
  F^g_{\mu\nu}(x) 
=& \kappa  ( \partial_\mu  {\rm tr}\{\bm{m}(x)\mathscr{A}_\nu(x) \} - \partial_\nu  {\rm tr}\{\bm{m}(x)\mathscr{A}_\mu(x) \}
+ ig_{{}_{\rm YM}}^{-1}  {\rm tr}\{ \bm{m}(x) [\partial_\mu  \bm{n}_k(x) , \partial_\nu  \bm{n}_k(x) ] \}  )
 ,
\nonumber\\ 
%  F^{g A}_{\mu\nu}(x) 
=&  \partial_\mu   \{\bm{m}^A(x)\mathscr{A}_\nu^A(x) \} - \partial_\nu   \{\bm{m}^A(x)\mathscr{A}_\mu^A(x) \}
- g_{{}_{\rm YM}}^{-1} f^{ABC}  \bm{m}^A(x)  \partial_\mu  \bm{n}_k^B(x) \partial_\nu  \bm{n}_k^C(x)  
 .
\label{Fg5}
\end{align}

Thus, we have arrived at the final form of the NAST for $SU(N)$ in arbitrary representation:
%\textbf{Theorem}
\begin{align}
 W_C[\mathscr{A}]  =& \int [d\mu(g)]_{\Sigma}
\exp \left[ -ig_{{}_{\rm YM}} \int_{\Sigma: \partial \Sigma=C} F^g  \right] , \quad
F^g := \frac12 f^g_{\mu\nu}(x) dx^\mu \wedge dx^\nu ,
%\quad
\nonumber\\ 
  F^g_{\mu\nu}(x) 
%=& \kappa  ( \partial_\mu  {\rm tr}\{\bm{m}(x)\mathscr{A}_\nu(x) \} - \partial_\nu  {\rm tr}\{\bm{m}(x)\mathscr{A}_\mu(x) \}
%+ ig_{{}_{\rm YM}}  {\rm tr}\{ \bm{m}(x) [\partial_\mu  \bm{n}_k(x) , \partial_\nu  \bm{n}_k(x) ] \}  ) ,
%\nonumber\\ 
%  F^{g A}_{\mu\nu}(x) 
=&  \Lambda_j \{  \partial_\mu   [ {n}_j^A(x)\mathscr{A}_\nu^A(x) ] - \partial_\nu   [ {n}_j^A(x)\mathscr{A}_\mu^A(x) ]
- g_{{}_{\rm YM}}^{-1} f^{ABC}   {n}_j^A(x)  \partial_\mu   {n}_k^B(x) \partial_\nu  {n}_k^C(x) \}
 ,
\nonumber\\ 
%    {m}^A(x) =& \Lambda_j  {n}_j^A(x) , \quad 
\bm{n}_j(x) =& g(x) H_j g^\dagger(x) = n^A_j(x) T_A \quad (j=1,...,r) . 
\label{Fg4}
\end{align}
 We can introduce also the  normalized  
\footnote{
This color field is normalized in the fundamental representation. 
In general, 
$2{\rm tr}[\bm{m}(x) \bm{m}(x)]
=\Lambda_j \Lambda_k 2{\rm tr}[\bm{n}_j(x) \bm{n}_k(x)]
=\Lambda_j \Lambda_k 2{\rm tr}[H_j H_k]
=\Lambda_j^2
$, which is equal to $\frac{N-1}{2N}$ in the fundamental representation.
}
and traceless field $\bm{n}(x)$ which we call the \textbf{color (direction) field} \cite{Kondo08}:
\begin{equation}
 \bm{n}(x) :=   \sqrt{\frac{2N}{ N-1 }}  {\bm{m}}(x) , \quad \text{or} \quad
%= \sqrt{\frac{N}{2(N-1)}} g(x) \left[ \rho - \frac{\bm{1}}{{\rm tr}(\bm{1})} \right]  g^\dagger(x)  
 \bm{m}(x) :=   \sqrt{\frac{ N-1 }{2N}}  {\bm{n}}(x)
 ,
 \label{C29-n-def}
\end{equation}
to rewrite the NAST into
\begin{align}
 W_C[\mathscr{A}]  =& \int [d\mu(g)]_{\Sigma}
\exp \left[ -ig_{{}_{\rm YM}} \sqrt{\frac{ N-1 }{2N}} \int_{\Sigma: \partial \Sigma=C} f^g  \right] , \quad
f^g := \frac12 f^g_{\mu\nu}(x) dx^\mu \wedge dx^\nu ,
%\quad
\nonumber\\ 
  f^g_{\mu\nu}(x) 
=& \kappa  ( \partial_\mu  {\rm tr}\{\bm{n}(x)\mathscr{A}_\nu(x) \} - \partial_\nu  {\rm tr}\{\bm{n}(x)\mathscr{A}_\mu(x) \}
+ ig_{{}_{\rm YM}}^{-1}  {\rm tr}\{ \bm{n}(x) [\partial_\mu  \bm{n}_k(x) , \partial_\nu  \bm{n}_k(x) ] \}  )
 ,
\nonumber\\ 
%  f^{g A}_{\mu\nu}(x) 
=&  \partial_\mu   \{ {n}^A(x)\mathscr{A}_\nu^A(x) \} - \partial_\nu   \{ {n}^A(x)\mathscr{A}_\mu^A(x) \}
- g_{{}_{\rm YM}}^{-1} f^{ABC}   {n}^A(x)  \partial_\mu   {n}_k^B(x) \partial_\nu   {n}_k^C(x)  
 ,
\nonumber\\ 
   \bm{n}(x) =& \sqrt{\frac{2N}{ N-1 }}  \Lambda_j \bm{n}_j(x) , \quad 
\bm{n}_j(x) = g(x) H_j g^\dagger(x) \quad (j=1,...,r) . 
\label{Fg4b}
\end{align}

In what follows, we work out the $G=SU(3)$ case for concreteness. 
For $G=SU(3)$, we choose the highest-weight state as the reference state. 
Then the highest-weight vector of the representation with the \textbf{Dynkin indices} $[m,n]$ is given by 
\begin{equation}
 \vec{\Lambda} 
 = (\Lambda_3,\Lambda_8) 
=  \left( \frac{m}{2}, \frac{m+2n}{2\sqrt{3}} \right) 
 .
\end{equation}
The fields $\bm{n}_3$ and $\bm{n}_8$ are independent of the representation and, hence, can be calculated in the fundamental representation:
\begin{align}
\bm{n}_3(x) = g(x) H_3 g^\dagger(x) = g(x) \frac{\lambda_3}{2} g^\dagger(x) , \quad
\bm{n}_8(x) = g(x) H_8 g^\dagger(x) = g(x) \frac{\lambda_8}{2} g^\dagger(x) ,
\end{align}
with the components:
\begin{align}
 {n}_3^A(x) = 2{\rm tr}\left[ \frac{\lambda_A}{2} g(x) \frac{\lambda_3}{2} g^\dagger(x) \right] , \quad
 {n}_8^A(x) = 2{\rm tr}\left[ \frac{\lambda_A}{2} g(x) \frac{\lambda_8}{2} g^\dagger(x) \right] ,
\end{align}
where $\lambda_3$ and $\lambda_8$ are the diagonal matrices of the Gell-Mann matrices $\lambda_A$ ($A=1,...,8$) for the Lie algebra $su(3)=Lie(SU(3))$. 
The parametrization of a group element $g$ and the explicit form of the integration measure $d\mu(g)$ can be found in \cite{KT00}.

%%%%%%%%%%%%%%%%%%%%%%%%%%%%%%%%%%%%%%%%%%%%%%%%%%%%%%%%%
\begin{figure}[ptb]
\begin{center}
\includegraphics[scale=0.25]{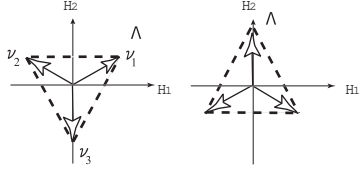}
\end{center}
\vskip -0.5cm
\caption{
The weight diagram for the fundamental representation of $SU(3)$,
(Left) $[1,0]={\bf 3}$,  where  $\vec{\Lambda} = \vec h_1=\vec \nu_1:=(\frac{1}{2},\frac{1}{ 2\sqrt{3}})$ is the highest weight and the other weights are 
$\vec\nu_2:=(-\frac{1}{2},\frac{1}{2\sqrt{3}})$and  $\vec\nu_3:=(0,-\frac{1}{\sqrt{3}})$,
 (Right) $[0,1]={\bf 3}^*$, the highest weight is  $-\vec\nu_3:=(0,\frac{1}{\sqrt{3}})$ and the other weights are 
$-\vec\nu_2:=(\frac{1}{2},-\frac{1}{2\sqrt{3}})$ and
$-\vec \nu_1:=(-\frac{1}{2},-\frac{1}{ 2\sqrt{3}})$.
}
 \label{C28-fig:fundamental-weight}
\end{figure}
%%%%%%%%%%%%%%%%%%%%%%%%%%%%%%%%%%%%%%%%%%%%%%%%%%%%%%%%%

%%%%%%%%%%%%%%%%%%%%%%%%%%%%%%%%%%%%%%%%%%%%%%%%%%%%%%%%%
\begin{figure}
\begin{center}
% \leavevmode
% \epsfxsize=60mm
% \epsfysize=60mm
% \epsfbox{root_SU3rev.eps}
\includegraphics[scale=0.25]{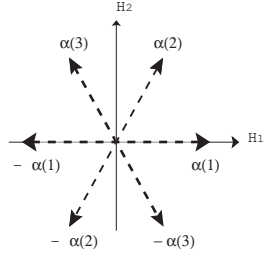}
\end{center} 
\vskip -0.5cm
\caption[]{
The root diagram of $SU(3)$ is equal to the weight diagram of the adjoint representation  $[1,1]={\bf 8}$ of $SU(3)$.
Here the positive root vectors are given by 
$\vec{\alpha}^{(1)}=(1,0)$,
$\vec{\alpha}^{(2)}=(\frac{1}{2},\frac{\sqrt{3}}{2})$, and
$\vec{\alpha}^{(3)}=(\frac{-1}{2},\frac{\sqrt{3}}{2})$.
%Here we have used the same weight ordering as in the $SU(N)$ case (see (\ref{C28-order}))  in defining the simple roots. Then 
The two simple roots are given by $\alpha^1:=
  \vec{\alpha}^{(1)} 
$
and
$ \alpha^2:= \vec{\alpha}^{(3)}$.
$\vec{\Lambda}=(\frac{1}{2},\frac{\sqrt{3}}{2})$ is the highest weight of the adjoint
representation.
}
 \label{C28-fig:root}
\end{figure}%%%%%%%%%%%%%%%%%%%%%%%%%%%%%%%%%%%%%%%%%%%%%%%%%%%%%%%%%

For the fundamental representation $[0,1]$, the color field takes the value in the Lie algebra of $SU(3)/U(2)=CP^2$ (See Fig.~\ref{C28-fig:fundamental-weight}):
\begin{align}
   \bm{m}(x) 
= \frac{1}{\sqrt{3}} \bm{n}(x) 
=   \frac{1}{\sqrt{3}} \bm{n}_8(x) 
= \frac{1}{\sqrt{3}} g(x) \frac{\lambda_8}{2} g^\dagger(x) 
= \frac{1}{6} g(x) 
\small
\begin{pmatrix}
  1 & 0 & 0 \cr
  0 & 1 & 0 \cr
  0 & 0 & -2 
 \end{pmatrix}
 g^\dagger(x)  \in Lie[SU(3)/U(2)]
 .
\end{align}
This is also the case for the fundamental representation $[1,0]$:
\begin{align}
   \bm{m}(x) 
=  \frac{1}{2} \bm{n}_3(x) + \frac{1}{2\sqrt{3}} \bm{n}_8(x)  
=  g(x) \left[ \frac{1}{2} \frac{\lambda_3}{2} + \frac{1}{2\sqrt{3}} \frac{\lambda_8}{2} \right] g^\dagger(x)
= \frac{-1}{6} g(x) 
\small
\begin{pmatrix}
  -2 & 0 & 0 \cr
  0 & 1 & 0 \cr
  0 & 0 & 1 
 \end{pmatrix}
 g^\dagger(x) \in Lie[SU(3)/U(2)]
 .
\end{align}
The fundamental representations have the same structure characterized by the degenerate matrix: the two of the three diagonal elements are equal, despite their different appearance. 

For the adjoint representation $[1,1]$, on the other hand, the color field takes the value in the Lie algebra of $SU(3)/U(1)^2=F^2$ (see Fig.~\ref{C28-fig:root}):
\begin{align}
   \bm{m}(x) 
=  \frac{1}{2} \bm{n}_3(x) + \frac{\sqrt{3}}{2} \bm{n}_8(x)  
=  g(x) \left[ \frac{1}{2} \frac{\lambda_3}{2} + \frac{\sqrt{3}}{2} \frac{\lambda_8}{2} \right] g^\dagger(x)
= \frac{1}{2} g(x) 
\small
\begin{pmatrix}
  1 & 0 & 0 \cr
  0 & 0 & 0 \cr
  0 & 0 & -1 
 \end{pmatrix}
 g^\dagger(x) \in Lie[SU(3)/U(1)^2]
 .
\end{align}
Here the matrix between $g$ and $g^\dagger$ is not degenerate: the three diagonal elements take different values.

For the general representation with the Dynkin index $[m,n]$, the color field reads
\begin{align}
   \bm{m}(x) = \frac{1}{\sqrt{3}} \bm{n}(x) 
=  \frac{m}{2} \bm{n}_3(x) + \frac{m+2n}{2\sqrt{3}} \bm{n}_8(x)  
%=  g(x) \left[ \frac{m}{2} \frac{\lambda_3}{2} + \frac{m+2n}{2\sqrt{3}} \frac{\lambda_8}{2} \right] g^\dagger(x)
= \frac{1}{3} g(x) 
\small
\begin{pmatrix}
  2m+n & 0 & 0 \cr
  0 & -m+n & 0 \cr
  0 & 0 & -m-2n 
 \end{pmatrix}
 g^\dagger(x) \in Lie[SU(3)/\tilde H]
 .
\end{align}
where $\tilde H$ is called the \textbf{maximal stability subgroup}.

%%%%%%%%%%%%%%%%%%%%%%%%%%%%%%%%%%%%%%%%%%%%%%%%%%%%%%%%%%%
%%%%%%%%%%%%%%%%%%%%%%%%%%%%%%%%%%%%%%%%%%%%%%%%%%%%%%%%%%%
\begin{figure}[ptb]
\begin{center}
\includegraphics[scale=0.25]{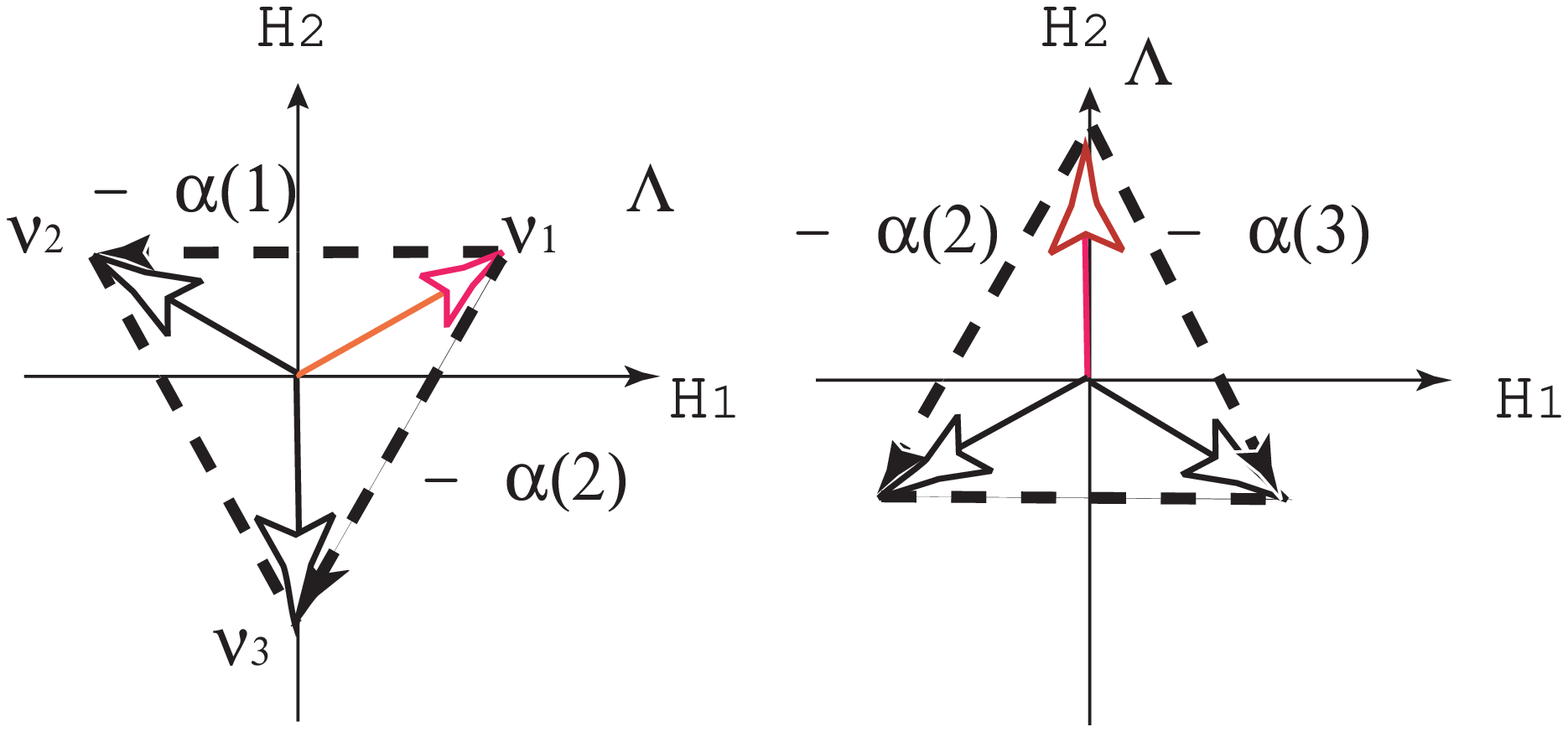}
\includegraphics[scale=0.25]{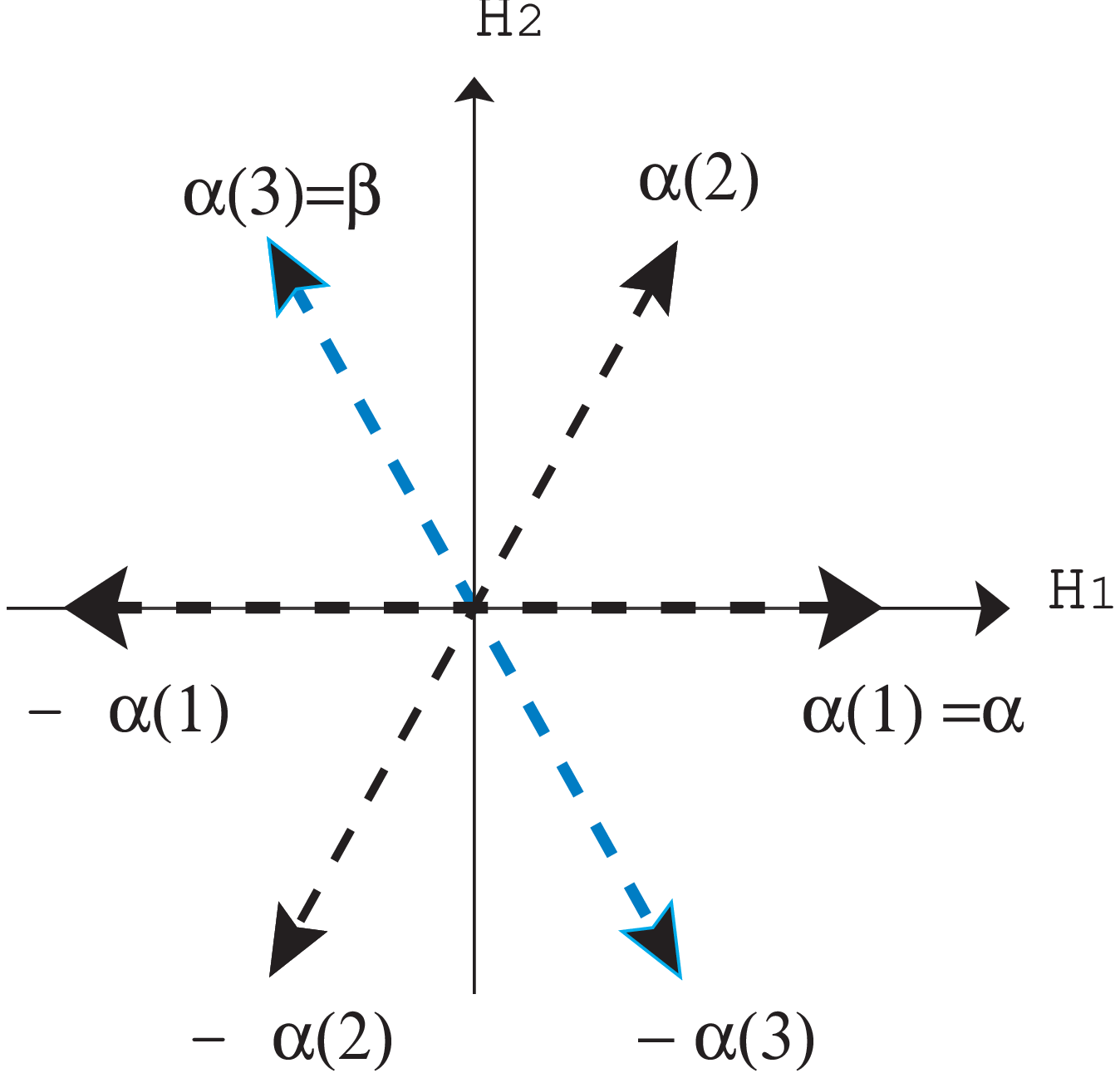}
\end{center}
\vskip -0.5cm
\caption{
The relationships among the weight vectors $\vec{\nu}_1, \vec{\nu}_2, \vec{\nu}_3$ in the fundamental representations ${\bf 3}$     and the root vectors $\vec{\alpha}^{(1)}, \vec{\alpha}^{(2)}, \vec{\alpha}^{(3)}$  in $SU(3)$.
We find
$
\vec{\nu}_1  \perp  \vec{\alpha}^{(3)}, - \vec{\alpha}^{(3)}
$.
Here   $\vec\Lambda  =\vec{\nu}_1:=(\frac{1}{2},\frac{1}{2\sqrt{3}})$ is the highest weight of the fundamental representation ${\bf 3}$.
}
 \label{C28-fig:fundamental-weight3}
\end{figure}
%%%%%%%%%%%%%%%%%%%%%%%%%%%%%%%%%%%%%%%%%%%%%%%%%%%%%%%%%%%
%%%%%%%%%%%%%%%%%%%%%%%%%%%%%%%%%%%%%%%%%%%%%%%%%%%%%%%%%%%

%%%%%%%%%%%%%%%%%%%%%%%%%%%%%%%%%%%%%%%%%%%%%%%%%%%%%%%%%
\begin{figure}
\begin{center}
% \leavevmode
% \epsfxsize=60mm
% \epsfysize=60mm
% \epsfbox{weight-adjointrev.eps}
\includegraphics[scale=0.25]{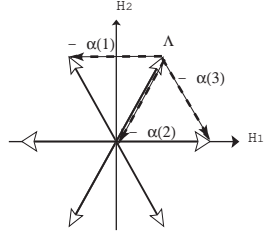}
\end{center} 
\vskip -0.5cm
\caption[]{
The weight vectors and root vectors required to define the coherent state in the adjoint representation $[1,1]={\bf 8}$ of $SU(3)$, where  $\vec{\Lambda}=(\frac{1}{2},\frac{\sqrt{3}}{2})$ is the highest weight of the adjoint
representation.
}
 \label{C28-fig:adjoint-weight}
\end{figure}
%%%%%%%%%%%%%%%%%%%%%%%%%%%%%%%%%%%%%%%%%%%%%%%%%%%%%%%%%

Thus,we can show that every representation $R$ of $SU(3)$ specified by the Dynkin index $[m,n]$ belongs to (I) or (II):
\begin{enumerate}
\item[(I)]
Minimal case: 
If $mn=0$  ($m=0$ or $n=0$), the maximal stability group
$\tilde H$ is given by 
\begin{equation}
 \tilde H=U(2) ,
\end{equation}
with generators 
$\{ H_1, H_2, E_\beta, E_{-\beta} \}$.
In the {minimal} case,  ${\rm dim}(G/\tilde{H})$ is minimal. 
Such a degenerate case occurs when the highest-weight vector $\vec{\Lambda}$ is orthogonal to some root vectors. %
%\footnote{
%The orthogonality of the highest-weight vector $\vec{\Lambda}$  to the respective root vector is realized as
%\begin{align}
% 0 =& \vec{\alpha}^{(1)} \cdot \vec{\Lambda} = \frac{m}{2} 
% \Longrightarrow m=0 \Longrightarrow [0,n] ,
% \nonumber\\
% 0 =& \vec{\alpha}^{(2)} \cdot \vec{\Lambda} = \frac{m+n}{2} 
% \Longrightarrow m=-n \Longrightarrow [m,-m] ,
% \nonumber\\
% 0 =& \vec{\alpha}^{(3)} \cdot \vec{\Lambda} = \frac{n}{2} 
% \Longrightarrow n=0 \Longrightarrow [m,0] ,
%\end{align}
%}   
In the minimal case, the coset $G/\tilde H$ is given by the \textbf{complex projective space}:
\begin{equation}
  G/\tilde H= SU(3)/U(2) =SU(3)/(SU(2)\times U(1))=CP^2,
\end{equation}
For example, the fundamental representation $[1,0]$ has the maximal stability subgroup $U(2)$ with the generators  
$\{ H_1, H_2, E_{\alpha^{(3)}}, E_{-\alpha^{(3)}} \} \in u(2)$,
where% $\Lambda \perp \beta, -\beta$. 
\begin{equation}
 \vec{\Lambda}=\vec{\nu}_1  \perp  \vec{\alpha}^{(3)}, - \vec{\alpha}^{(3)}  . 
\end{equation}
See Fig.~\ref{C28-fig:fundamental-weight3}.

\item[(II)]
Maximal case:
If $mn \not=0$ 
($m\not=0$ and $n\not=0$), $\tilde H$ is the maximal torus group: 
\begin{equation}
\tilde H=H=U(1) \times U(1) ,
\end{equation}
with generators 
$\{ H_1, H_2 \}$.
In the {maximal} case,  ${\rm dim}(G/\tilde{H})$ is maximal. 
This is a non-degenerate case.
In the maximal case, the coset $G/\tilde H$ is given by the \textbf{flag space}:
\begin{equation}
  G/\tilde H=  SU(3)/(U(1)\times U(1))=F_2 .
\end{equation}
For example, the adjoint representation $[1,1]$ has the maximal stability subgroup $U(1) \times U(1)$ with the generators $\{ H_1, H_2  \} \in u(1)+u(1)$.
See Fig.\ref{C28-fig:adjoint-weight}.
\end{enumerate}

%In the minimal case, the coset $G/\tilde H$ is given by the \textbf{complex projective space}
%\begin{equation}
%  G/\tilde H= SU(3)/U(2) =SU(3)/(SU(2)\times U(1))=CP^2,
%\end{equation}
%whereas in the maximal case, the coset $G/\tilde H$ is given by the \textbf{flag space}
%\begin{equation}
%  G/\tilde H=  SU(3)/(U(1)\times U(1))=F_2 .
%\end{equation}
%Here, $CP^{n}$ is the \textbf{complex projective space} and $F_n$ is the \textbf{flag space}. 
%Therefore, \textit{the   fundamental representations ($\bm{3}$ and $\bm{3^*}$) of $SU(3)$ belong to the minimal case (I), and hence the maximal stability group is $U(2)$, rather than the maximal torus group $U(1) \times U(1)$}.  

%%%%%%%%%%%%%%%%%%%%%%%%%%%%%%%%%%%%%%%%%%%%%%%%%

\section{Magnetic monopoles}

%%%%%%%%%%%%%%%%%%%%%%%%%%%%%%%%%%%%%%%%%%%%%%%%

We can define the gauge-invariant \textbf{magnetic-monopole} current 
  $k= \delta {}^*f^g$
%  $k=\frac{1}{\sqrt{3}}\delta {}^*\mathscr{G}$
from the field strength $f^g_{\mu\nu}$ through the NAST. 
% (in the vector notation):
%\begin{align}
% \mathscr{G}_{\mu\nu}  
%  :=   \partial_\mu 2{\rm tr} \{ \bm{n}  \mathscr{A}_\nu  \} - \partial_\nu 2{\rm tr} \{ \bm{n}  \mathscr{A}_\mu  \} 
%\nonumber\\& 
%+ \frac{4}{3} ig_{{}_{\rm YM}}^{-1} 2{\rm tr} \{ \bm{n} [\partial_\mu \bm{n} , \partial_\nu \bm{n}  ] \} 
% .
%\end{align}
The magnetic--monopole current $k$ is defined as the ($D-3$)-form using the gauge-invariant field strength (curvature two-form)  by
\begin{equation}
 k  = \delta {}^{\displaystyle *}
 f^g = {}^{\displaystyle *}df^g , 
 \quad 
 f^g = \sum_{j=1}^{r} \Lambda_j f^{(j)} 
   .
\end{equation}
Using the same procedure as given in \cite{Kondo08}, the Wilson loop operator in arbitrary representation of $SU(N)$ is written in terms of the electric current $j$ and the magnetic current $k$:
\begin{equation}
 W_C[\mathscr{A}] 
= \int [d\mu({g})] \exp \left\{ -ig_{{}_{\rm YM}}    
%\frac12 \sqrt{\frac{2(N-1)}{N}} 
  \sqrt{\frac{ N-1 }{2N}} [ (  \omega_{\Sigma_C}, k) +    (  N_{\Sigma_C},j  ) ] \right\} ,
\end{equation}
where we have defined the  $(D-3)$-form  $k$ and the one-form $j$ in $D$ spacetime dimensions:
\begin{align}
k:=  \delta  {}^{\displaystyle *}f^g 
%=  {}^{\displaystyle *}df^g 
, \quad
j:=& \delta f^g , 
%\quad
%G  := 2{\rm tr}\{ \bm{n} \mathscr{F} [\mathscr{V}]\} .
\end{align}
we have introduced an antisymmetric tensor $\Theta_{\Sigma_C}$ of rank two which has the support only on the surface $\Sigma_C$ spanned by the  loop $C$:
%with the surface element $dS^{\mu\nu}$ of $\Sigma_C$ spanned by the  loop $C$: 
\begin{align}
 \Theta^{\mu\nu}_{\Sigma_C} (x) 
:=   \int_{\Sigma_C: \partial \Sigma_C=C}  d^2 S^{\mu\nu}(x(\sigma)) \delta^D(x-x(\sigma)) 
%=  {1 \over 2} \int_{\Sigma_C} d^2 \sigma J^{\mu\nu}(\sigma) \delta^D(x-x(\sigma)) 
 ,
%=  - \Theta_{\nu\mu}(x) .
\end{align}
and
%we have defined the $(D-3)$-form $k$ and one-form $j$ by
%\begin{align}
%k:=  \delta *G  , \quad  j:=& \delta G  , \quad
%G_{\mu\nu} := {\rm tr}\{ \bm{n} \mathscr{F}_{\mu\nu} [\mathscr{V}]\} ,
%\end{align}
%and
we have defined the $(D-3)$-form $\omega_{\Sigma_C}$ and one-form $N_{\Sigma_C}$  using the Laplacian $\Delta$ by
\begin{equation}
 \omega_{\Sigma_C} :=  {}^{\displaystyle *} d \Delta^{-1} \Theta_{\Sigma_C}  = \delta  \Delta^{-1}  {}^{\displaystyle *}\Theta_{\Sigma_C}  , \quad
 N_{\Sigma_C} := \delta  \Delta^{-1}  \Theta_{\Sigma_C} ,
\end{equation}
with the inner product for two forms being defined by
\begin{align}
 & ( \omega_{\Sigma_C},k) 
= \frac{1}{(D-3)!} \int d^Dx k^{\mu_1 \cdots \mu_{D-3}}(x) \omega^{\mu_1 \cdots \mu_{D-3}}_{\Sigma_C}(x) ,
  \nonumber\\
 &
  (N_{\Sigma_C},j) =  \int d^Dx j^{\mu}(x) N^{\mu}_{\Sigma_C}(x) .
\end{align}
%We call $\Theta^{\mu\nu}_{\Sigma_C}(x)$ the \textbf{vorticity tensor} which has   the support only on the surface $\Sigma_C$ spanned by the  loop $C$.
Here we have replaced the measure $[d\mu({g})]_{\Sigma}$ by $[d\mu({g})]:=[d\mu({g})]_{\mathbb{R}^D}=\prod_{x \in \mathbb{R}^D} d\mu({g}(x))$ over all the spacetime points.

%First, we consider a special case in which $U(x)$ is rotated to the identity matrix by performing a gauge transformation so that the one-form $V$ takes the form:
%\begin{equation}
% V(x) \rightarrow  {\rm tr}(  \mathcal{H} \mathscr{A}(x)) 
% ,
%\end{equation}
%which reads 
%for $[1,0]$ and $[-1,0]$,  
%\begin{equation}
%  V(x)   \rightarrow  \pm   \left[ \frac{1}{2} \mathscr{A}^3(x) +   \frac{1}{2\sqrt{3}} \mathscr{A}^8(x) \right]
%  =  \pm  (\mathscr{A}^A(x)T_A)_{11}
%  , 
%\end{equation}
%and  $[0,-1]$ and $[0,1]$, 
%\begin{equation}
%  V(x)   \rightarrow  \pm   \left[ -\frac{1}{2} \mathscr{A}^3(x)  +   \frac{1}{2\sqrt{3}} \mathscr{A}^8(x) \right] 
%  =  \pm  (\mathscr{A}^A(x)T_A)_{22}
%  , 
%\end{equation}
%In particular, for $[-1,1]$ and $[1,-1]$,  
%\begin{equation}
% V(x)   \rightarrow   \pm   \left[  \frac{-2}{2\sqrt{3}} \mathscr{A}^8(x) \right]
%  =  \pm   (\mathscr{A}^A(x)T_A)_{33}
%   .
%\end{equation}
For $D=4$, especially, the magnetic current reads 
\begin{equation}
 k^\mu = \frac12 \epsilon^{\mu\nu\rho\sigma} \partial_\nu f_{\rho\sigma}^g ,
 \quad 
 f_{\mu\nu}^g = \sum_{j=1}^{r} \Lambda_j f^{(j)}_{\mu\nu}  
%\Lambda_1 f_{\mu\nu}^{(1)} + \Lambda_2 f_{\mu\nu}^{(2)}
   .
\end{equation}
Then, the \textbf{magnetic charge} is defined by  
\begin{equation}
  q_m = \int d^3x k^0 
= \int d^3x \frac12 \epsilon^{jk\ell} \partial_\ell    f_{jk}^g(x) 
= \int d^2S_\ell  \epsilon^{jk\ell} \frac12   f_{jk}^g(x) 
%= \int d^2S^{jk}    F_{jk}(x) 
   .
   \label{C29-m-charge}
\end{equation}
We examine the \textbf{quantization condition} for the magnetic charge.
In the $SU(3)$ case, the two kinds of gauge-invariant field strength are given by 
\begin{align}
 f_{\mu\nu}^g =& \Lambda_1 f_{\mu\nu}^{(1)} + \Lambda_2 f_{\mu\nu}^{(2)} ,
\nonumber\\
 f_{\mu\nu}^{(1)}  
%:=& \bm{n}_3 \cdot \mathscr{F}_{\mu\nu}[\mathscr{V}]
=&  \partial_\mu  2{\rm tr}\{ \bm{n}_3 \mathscr{A}_\nu  \} - \partial_\nu 2{\rm tr}\{ \bm{n}_3 \mathscr{A}_\mu  \}
%\nonumber\\& 
-  ig_{{}_{\rm YM}}^{-1} 2{\rm tr}\{  \bm{n}_3  [ \partial_\mu \bm{n}_3  , \partial_\nu \bm{n}_3 ]  + \bm{n}_3 [ \partial_\mu \bm{n}_8  , \partial_\nu \bm{n}_8 ] \} ,
\nonumber\\
 f_{\mu\nu}^{(2)}  
%:=& \bm{n}_8 \cdot \mathscr{F}_{\mu\nu}[\mathscr{V}]
  =&   \partial_\mu  2{\rm tr}\{ \bm{n}_8    \mathscr{A}_\nu  \} - \partial_\nu 2{\rm tr}\{ \bm{n}_8  \mathscr{A}_\mu  \}
%\nonumber\\& 
-  ig_{{}_{\rm YM}}^{-1} 2{\rm tr}\{  \bm{n}_8  [ \partial_\mu \bm{n}_3  , \partial_\nu \bm{n}_3 ]  + \bm{n}_8 [ \partial_\mu \bm{n}_8  , \partial_\nu \bm{n}_8 ] \} 
\nonumber\\  
  =&   \partial_\mu  2{\rm tr}\{ \bm{n}_8    \mathscr{A}_\nu  \} - \partial_\nu 2{\rm tr}\{ \bm{n}_8  \mathscr{A}_\mu  \}
%\nonumber\\& 
- \frac{4}{3} ig_{{}_{\rm YM}}^{-1} 2{\rm tr}\{   \bm{n}_8 [\partial_\mu \bm{n}_8  , \partial_\nu \bm{n}_8 ]) 
 .
\label{f1f2} 
\end{align}
Notice that $f_{\mu\nu}^{(2)}$ is written in terms of $\bm{n}_8$ alone  
(see Appendix~\ref{section:monopole-fs} for the derivation of $f_{\mu\nu}^{(2)}$).
It is shown \cite{KKSS15} that the two kinds of the   gauge-invariant charges $q_m^{(1)}$ and $q_m^{(2)}$ obey the different quantization conditions:% \cite{GNO77}:%
%\footnote{
%See, e.g., % EW(1976). GNO (1977).
%\bibitem{EW76}
%E. Weinberg, 
%\bibitem{GNO77}
%P. Goddard, J. Nuyts, David I. Olive,
%Gauge Theories and Magnetic Charge, 
%Nucl. Phys. B{\bf 125}, 1--28 (1977). 
%\cite{EW76,GNO77}:
%}
\begin{align} 
q_m =& \Lambda_1 q_m^{(1)} + \Lambda_2 q_m^{(2)} ,
\nonumber\\
 q_m^{(1)} 
:=& \int d^3x \frac12 \epsilon^{jk\ell} \partial_\ell f_{jk}^{(1)}(x) 
%({\bm n}_1(x) , \mathscr{F}_{jk}[\mathscr{V}](x))
= \frac{4\pi}{g_{{}_{\rm YM}}} \left( n- \frac12 n' \right) , 
\nonumber\\
 q_m^{(2)} 
:=& \int d^3x \frac12 \epsilon^{jk\ell} \partial_\ell f_{jk}^{(2)}(x) 
%({\bm n}_2(x) , \mathscr{F}_{jk}[\mathscr{V}](x))
= \frac{4\pi}{g_{{}_{\rm YM}}} \frac12 \sqrt{3} n' , \quad n, n' \in \mathbb{Z} 
  .
\label{C29-qc2}
\end{align}
The existence of the magnetic charge $q_m^{(1)}$ characterized by two integers $n$ and $n'$ is consistent with a fact that 
the map defined by
\begin{equation}
   \bm{n}_3: S^2 \rightarrow SU(3)/[U(1)\times U(1)] \simeq F_2 ,
\end{equation}
 has the nontrivial Homotopy group:  
\begin{equation}
   \pi_2(SU(3)/[U(1) \times U(1)])=\pi_1(U(1) \times U(1)) 
 =\mathbb{Z} + \mathbb{Z} .
\end{equation}
On the other hand, the existence of the magnetic charge $q_m^{(2)}$ characterized by an integer   $n'$ is consistent with a fact that 
the map defined by
\begin{equation}
   \bm{n}_8: S^2 \rightarrow SU(3)/U(2) \simeq \mathbb{C}P^2   ,
\end{equation}
 has the following nontrivial homotopy group:  
\begin{equation}
   \pi_2(SU(3)/[SU(2) \times U(1)])=\pi_1(SU(2) \times U(1)) 
%\nonumber\\&
=\pi_1(U(1))=\mathbb{Z} .
\end{equation}
%The magnetic charge defined by $Q_m := \int d^3x k^0$ obeys the quantization condition characterized by an integer $\ell$: 
%\begin{equation}
% Q_m := \int d^3x k^0 = 2\pi \sqrt{3} g_{{}_{\rm YM}}^{-1} \ell , \ \ell \in \mathbb{Z} .
%\end{equation}
 
%In order to extract the information on the magnetic charge through the non-Abelian Stokes theorem for the Wilson loop operator, we consider the integration of the field strength $F=dV$ (the curvature two-form) over the closed surface $\Sigma$.   

%For the fundamental representation, $F=dV$  agrees with the diagonal components of the non-Abelian field $ \mathscr{A}(x)=\mathscr{A}^A(x)T_A$. 
%This fact is also seen from 
%\begin{equation}
% V(x)   \rightarrow 2{\rm tr}(\bm{m}(x) \mathscr{A}(x))  
%=  m^A(x)  \mathscr{A}^A(x)
%=  \left< \bm{\Lambda} |  G^{\dagger}(x)  \mathscr{A}(x) G(x) |\bm{\Lambda} \right>
% =  \left< U(x), \Lambda |  \mathscr{A} (x)   |U(x), \Lambda \right>
%   ,
%\end{equation}
%where
%$| \bm{\Lambda} >=(1,0,0)$,
%$| \bm{\Lambda} >=(0,1,0)$, and
%$| \bm{\Lambda} >=(0,0,1)$ for three fundamental representations.  

Incidentally, we can show  \cite{KKSS15} that the gauge-invariant field strength $F^g_{\mu\nu}$ is equal to the component of the non-Abelian field strength $\mathscr{F}[\mathscr{V}]$ of the restricted field $\mathscr{V}$ (in the decomposition $\mathscr{A}=\mathscr{V}+\mathscr{X}$) projected to the color field $\bm{n}$:
\begin{align}
F^g_{\mu\nu} =  {\rm tr}\{ \bm{m} \mathscr{F}_{\mu\nu} [\mathscr{V}]\} = \Lambda_j f_{\mu\nu}^{(j)}, \quad
 f_{\mu\nu}^{(j)} =  {\rm tr}\{ \bm{n}_j \mathscr{F} [\mathscr{V}]\} .
%=  \bm{n}_j \cdot \mathscr{F}_{\mu\nu}[\mathscr{V}]
\end{align}
This relation is useful in calculating the contribution from magnetic monopoles to the Wilson loop average from the viewpoint of the dual superconductor picture for quark confinement. 
The results will be given elsewhere. 

%%%%%%%%%%%%%%%%%%%%%%%%%%%%%%%%%%%%%%%%%%%%%%%%%

%\section{Conclusion and discussion}

%%%%%%%%%%%%%%%%%%%%%%%%%%%%%%%%%%%%%%%%%%%%%%%%%

{\it Acknowledgements}\ ---

The authors would like to thank Toru Shinohara for discussions on the field strength for the magnetic monopole. 
This work is  supported by Grants-in-Aid for Scientific Research (C) No.24540252 and (C) No.15K05042 from the Japan Society for the Promotion of Science (JSPS).

\appendix
\section{Derivation of eq.(\ref{dnB}) and eq.(\ref{Fg3})}\label{section:derivation-eq}

Equation (\ref{dnB}) is derived as follows.
\begin{align}
 ig_{{}_{\rm YM}} [\mathscr{B}_\mu , \bm{n}_j  ]  
=& - [[ \bm{n}_k , \partial_\mu  \bm{n}_k  ] , \bm{n}_j ]  
\nonumber\\
=&   [ [\partial_\mu  \bm{n}_k , \bm{n}_j ] , \bm{n}_k   ]  + [   [\bm{n}_j   , \bm{n}_k ], \partial_\mu  \bm{n}_k   ]  
\nonumber\\
=&   [ [\partial_\mu  \bm{n}_j ,  \bm{n}_k] , \bm{n}_k   ]   
\nonumber\\
=&   [ \bm{n}_k, [\bm{n}_k , \partial_\mu  \bm{n}_j  ]  ]   
\nonumber\\
=& \partial_\mu \bm{n}_j -  \bm{n}_k (\bm{n}_k, \partial_\mu \bm{n}_j)  .
\label{Bn}
\end{align}
where we have used the Jacobi identity in the second equality, the relation 
$[ \partial_\mu \bm{n}_k , \bm{n}_j ] = [\partial_\mu \bm{n}_j , \bm{n}_k ]$ 
following from $\partial_\mu [\bm{n}_k , \bm{n}_j]=0$
and the commutativity $[\bm{n}_j, \bm{n}_k ]=0$ in the third equality, and the identity 
$
\mathscr{F}=\bm{n}_k (\bm{n}_k, \mathscr{F})+[ \bm{n}_k, [\bm{n}_k , \mathscr{F} ]  ]
$ 
(see e.g., Appendix C of \cite{KKSS15})
in the fifth equality.
Moreover, we find that the last term in (\ref{Bn}) vanishes:
\begin{align}
  (\bm{n}_k, \partial_\mu \bm{n}_j)   
=& \kappa {\rm tr}(\bm{n}_k \partial_\mu \bm{n}_j)
\nonumber\\
=& \kappa {\rm tr}(g H_k g^\dagger \partial_\mu (g H_j g^\dagger) )
\nonumber\\
=& \kappa {\rm tr}( g H_k g^\dagger \partial_\mu g H_j g^\dagger ) + \kappa {\rm tr}( g H_k g^\dagger g H_j \partial_\mu g^\dagger )
\nonumber\\
=& - \kappa {\rm tr}( g H_k \partial_\mu g^\dagger  g H_j g^\dagger ) + \kappa {\rm tr}( g H_k  H_j \partial_\mu g^\dagger )
\nonumber\\
=& - \kappa {\rm tr}( g H_j g^\dagger g H_k \partial_\mu g^\dagger   ) + \kappa {\rm tr}( g H_k  H_j \partial_\mu g^\dagger )
\nonumber\\
=& - \kappa {\rm tr}( g H_j H_k \partial_\mu g^\dagger   ) + \kappa {\rm tr}( g H_k  H_j \partial_\mu g^\dagger )
\nonumber\\
=&   \kappa {\rm tr}( g [H_k , H_j] \partial_\mu g^\dagger )
= 0  ,
\label{ndn}
\end{align}
where we have used $g^\dagger g=\bm{1}=g g^\dagger$ and $g^\dagger \partial_\mu g=-\partial_\mu  g^\dagger g$ following from $\partial_\mu(g g^\dagger )=0$ in the fourth equality and cyclicity of the trace in the fifth equality.
Combining (\ref{Bn}) and (\ref{ndn}), indeed, we have (\ref{dnB}).

Equation (\ref{Fg3}) is derived as follows. 
The third term in $F^g_{\mu\nu}(x)$ is rewritten using the relation (\ref{dmm}) as 
\begin{align}
  ig_{{}_{\rm YM}}   {\rm tr}(\bm{m} [\Omega_\mu , \Omega_\nu ]) 
=&  
  ig_{{}_{\rm YM}}   {\rm tr}([\bm{m},\Omega_\mu ] \Omega_\nu  ) 
\nonumber\\ 
=& ig_{{}_{\rm YM}}   {\rm tr}(\Omega_\nu [\bm{m}, \mathscr{B}_\mu ]   )  
\nonumber\\ 
=& ig_{{}_{\rm YM}}   {\rm tr}([\Omega_\nu , \bm{m}] \mathscr{B}_\mu )  
\nonumber\\ 
=& ig_{{}_{\rm YM}}   {\rm tr}([\mathscr{B}_\nu , \bm{m}] \mathscr{B}_\mu )  
\nonumber\\ 
=&  {\rm tr}(\partial_\nu \bm{m} \mathscr{B}_\mu )  
\nonumber\\ 
=& ig_{{}_{\rm YM}}^{-1}  {\rm tr}(\partial_\nu \bm{m} [ \bm{n}_j , \partial_\mu  \bm{n}_j ])  
\nonumber\\ 
=& ig_{{}_{\rm YM}}^{-1}  {\rm tr}( [\partial_\nu \bm{m} ,   \bm{n}_j ] \partial_\mu  \bm{n}_j  )  
\nonumber\\ 
=& ig_{{}_{\rm YM}}^{-1}  {\rm tr}( [ \partial_\nu  \bm{n}_j , \bm{m} ] \partial_\mu  \bm{n}_j  )  
\nonumber\\ 
=& ig_{{}_{\rm YM}}^{-1}  {\rm tr}( [\partial_\mu  \bm{n}_j , \partial_\nu  \bm{n}_j ] \bm{m} )  
,
%\label{Fg3}
\end{align} 
where we have used ${\rm tr}\{ A [ B, C ]\}={\rm tr}\{ [A , B]  C  \}={\rm tr}\{ B [ C , A ]\}={\rm tr}\{ C [ A , B ]\}$ due to the cyclicity of the trace in the first, third, and seventh equalities, and the relation
$[ \partial_\nu \bm{m} , \bm{n}_j ] = [\partial_\nu \bm{n}_j , \bm{m} ]$ 
which is derived from
$\partial_\nu [\bm{m} , \bm{n}_j]=0$
and the commutativity
$[ \bm{m} , \bm{n}_j ]= 0$ in the eighth equality.
%\begin{align}
%  [ \bm{m} , \bm{n}_j ]= 0 \Longrightarrow \partial_\nu [\bm{m} , \bm{n}_j]=0 \Longrightarrow [ \partial_\nu \bm{m} , \bm{n}_j ] = [\partial_\nu \bm{n}_j , \bm{m} ]. 
%\end{align}

\section{Field strength for the magnetic monopole}\label{section:monopole-fs}

For $G=SU(N)$, we can define three types of products: $\cdot$,  $\times$, and $*$ in the vector form by 
\begin{subequations}
\begin{align}
\mathbf X \cdot \mathbf Y
 :=& X^A Y^A = \mathbf Y \cdot \mathbf X ,%= 2 {\rm tr}(\mathscr X \mathscr Y),
\\
(\mathbf X\times\mathbf Y)^C  
 :=& f_{ABC}X^AY^B  = - (\mathbf Y \times\mathbf X)^C ,
%\quad 
% [\mathscr X , \mathscr Y ] :=  if_{ABC}X^AY^B T_C   ,
\\
(\mathbf X*\mathbf Y)^C :=& d_{ABC}X^AY^B = (\mathbf Y*\mathbf X)^C, 
%\quad 
%  \{ \mathscr X , \mathscr Y \} - \frac{1}{N} 2 {\rm tr}(\mathscr X \mathscr Y) \mathbf{1} = d_{ABC}X^AY^B T_C 
\end{align}
\end{subequations}
which correspond to three operations in the Lie algebra form: ${\rm tr}()$, $[, ]$, and $\{, \}$ as\begin{subequations}
\begin{align}
 2 {\rm tr}(\mathscr X \mathscr Y)  =& \mathscr  X^A \mathscr Y^A ,
\\
 [\mathscr X , \mathscr Y ] =&  if_{ABC} \mathscr X^A \mathscr Y^B T_C   ,
\\
  \{ \mathscr X , \mathscr Y \} - \frac{1}{N} 2 {\rm tr}(\mathscr X \mathscr Y) \mathbf{1} 
=& d_{ABC} \mathscr X^A \mathscr Y^B T_C 
 .
\end{align}
\end{subequations}
Then we obtain the relation:
\begin{align}
 \mathbf{n}_8 \cdot (\partial_\mu \mathbf{n}_8 \times \partial_\nu \mathbf{n}_8 )
 =& \partial_\nu \mathbf{n}_8 \cdot (\mathbf{n}_8   \times \partial_\mu \mathbf{n}_8) 
 \nonumber\\
 =& (2\sqrt{3} \mathbf{n}_3 * \partial_\nu \mathbf{n}_3) \cdot (\mathbf{n}_8   \times \partial_\mu \mathbf{n}_8) 
 \nonumber\\
 =&  2\sqrt{3} \partial_\nu \mathbf{n}_3  \cdot [\mathbf{n}_3 * (\mathbf{n}_8   \times \partial_\mu \mathbf{n}_8)    ]   
 \nonumber\\
 =&  2\sqrt{3} \partial_\nu \mathbf{n}_3  \cdot [ \frac{\sqrt{3}}{2} (\mathbf{n}_8   \times \partial_\mu \mathbf{n}_3)    ]   
 \nonumber\\
 =&  3 \mathbf{n}_8 \cdot ( \partial_\mu \mathbf{n}_3 \times \partial_\nu \mathbf{n}_3)     
  ,
\end{align}
where we have used the identity:
$
\bm{X} \cdot (\bm{Y} \times \bm{Z}) = \bm{Y} \cdot (\bm{Z} \times \bm{X}) = \bm{Z} \cdot (\bm{X} \times \bm{Y})
$
in the first and the fifth equalities, 
$
\sqrt{3} \mathbf{n}_3 * \mathbf{n}_3 = \mathbf{n}_8 
$
in the second equality, 
and
$
(\bm{X} * \bm{Y}) \cdot \bm{Z} = (\bm{X} * \bm{Z}) \cdot \bm{Y}
$
in the third equality.
The fourth equality is shown as
\begin{align}
 \mathbf{n}_3 * (\mathbf{n}_8   \times \partial_\mu \mathbf{n}_8)    
  =&  \mathbf{n}_8  \times (\mathbf{n}_3 * \partial_\mu \mathbf{n}_8)
 \nonumber\\
  =&  \mathbf{n}_8  \times (\partial_\mu (\mathbf{n}_3 * \mathbf{n}_8) - \partial_\mu \mathbf{n}_3 * \mathbf{n}_8 )
 \nonumber\\
  =& \frac{1}{\sqrt{3}} \mathbf{n}_8  \times \partial_\mu  \mathbf{n}_3  - \mathbf{n}_8  \times ( \mathbf{n}_8 * \partial_\mu \mathbf{n}_3  ) 
 \nonumber\\
  =& \frac{1}{\sqrt{3}} \mathbf{n}_8  \times \partial_\mu  \mathbf{n}_3  + \frac{1}{2\sqrt{3}} \mathbf{n}_8  \times   \partial_\mu \mathbf{n}_3 
 \nonumber\\
 =&   \frac{\sqrt{3}}{2} (\mathbf{n}_8   \times \partial_\mu \mathbf{n}_3)   
  .
\end{align}
where we have used
$
\bm{X} * (\bm{Y} \times \bm{Z}) = \bm{Y} \times (\bm{X} * \bm{Z}) + \bm{Z} * (\bm{X} \times \bm{Y})
$ 
and
$\mathbf{n}_3 \times \mathbf{n}_8= 0$
in the first equality, 
the Leibniz rule in the second equality, 
$
\sqrt{3} \mathbf{n}_3 * \mathbf{n}_8 = \mathbf{n}_3 
$
in the third equality,
and
$
\bm{X} \times (\bm{X} * \bm{Z}) = \frac12 (\bm{X} * \bm{X}) \times \bm{Z}
$
following from
$
\bm{X} \times (\bm{Y} * \bm{Z}) = (\bm{X} * \bm{Y}) \times \bm{Z} +  (\bm{X} * \bm{Z}) \times \bm{Y}
$ 
and
$
\sqrt{3} \mathbf{n}_8 * \mathbf{n}_8 = -\mathbf{n}_8 
$
in the fourth equality.

This relation was used to write $f_{\mu\nu}^{(2)}$ in the form given in (\ref{f1f2}): 
\begin{align}
 \mathbf{n}_8 \cdot ( \partial_\mu \mathbf{n}_3 \times \partial_\nu \mathbf{n}_3) + \mathbf{n}_8 \cdot (\partial_\mu \mathbf{n}_8 \times \partial_\nu \mathbf{n}_8 )
 =   \frac{4}{3} \mathbf{n}_8 \cdot ( \partial_\mu \mathbf{n}_8 \times \partial_\nu \mathbf{n}_8)     
 .
\end{align}

%%%%%%%%%%   REFERENCES   %%%%%%%%%%

\end{document}